# TED-net: Convolution-free T2T Vision Transformer-based Encoder-decoder Dilation network for Low-dose CT Denoising

Dayang Wang[1], Zhan Wu[1], Hengyong Yu[1,*]

Department of Electrical and Computer Engineering, University of Massachusetts Lowell, Lowell, MA, USA.
*. Corresponding author, email: Hengyong-yu@ieee.org

**Abstract.** Low dose computed tomography is a mainstream for clinical applications. However, compared to normal dose CT, in the low dose CT (LDCT) images, there are stronger noise and more artifacts which are obstacles for practical applications. In the last few years, convolution-based end-to-end deep learning methods have been widely used for LDCT image denoising. Recently, transformer has shown superior performance over convolution with more feature interactions. Yet its applications in LDCT denoising have not been fully cultivated. Here, we propose a convolution-free T2T vision transformer-based Encoder-decoder Dilation network (TED-net) to enrich the family of LDCT denoising algorithms. The model is free of convolution blocks and consists of a symmetric encoder-decoder block with sole transformer. Our model is evaluated on the AAPM-Mayo clinic LDCT Grand Challenge dataset, and results show outperformance over the state-of-the-art denoising methods.

**Keywords:** Low-dose CT, transformer, encoder decoder, dilation.

## 1 Introduction

In recent years, low dose computed tomography (LDCT) has become the mainstream in the clinical applications of medical imaging. However, the low quality of LDCT image has always been a barrier since it compromises the diagnosis value. To overcome this issue, traditional methods (e.g. iterative methods) manage to suppress the artifact and noise by using the physical model and/or prior information. For example, Compressive Sensing (CS) has been widely used for ill-posed inverse problems by learning sparse representations [1], and the representative total variation (TV)-based models assume that the clean image is piecewise constant and its gradient transform is sparse [2-5]. Xu *et al* combined dictionary learning and statistic IR (SIR) [6] for LDCT denoising. Tan *et al* proposed a tensor-based dictionary learning model for spectral and dynamic CT [7]. Ma *et al* designed a Non-Local Mean (NLM) method to utilize the redundancy of information across the whole image rather than local operations on neighboring image voxels [8]. Nonetheless, none of these algorithms is adopted in commercial scanners because of the hardware limitations and high computational cost [9].

In last few years, deep learning-based methods attracted more attention for CT denoising and achieved the state-of-the-art performance [10]. Chen *et al* combined the auto-encoder, deconvolution network, and shortcut connections into a residual encoder-decoder convolution neural network (CNN) for CT imaging [11]. Yang *et al* used WGAN-VGG in the denoising base network and adopted perceptual loss to evaluate the reconstructed image quality [12]. Fan *et al* constructed a quadratic neuron-based autoencoder with more robustness and utility for model efficiency in contradiction of other CT denoising methods [13]. However, these convolution-based methods has limited ability to capture contextual information with long spatial dependence in image or feature maps.

Very recently, transformer [14] has gradually become the dominant method in the natural language processing (NLP) [15-18] and computer vision (CV) fields [19-31]. Transformer has achieved a great performance in high level tasks, such as classification, object detection, image segmentation, *etc*. Dosovitskiy *et al* first proposed vision transformer (ViT) in the CV field by mapping an image into 16×16 sequence words [23]. To overcome the simple tokenization in ViT, Yuan *et al* further proposed a Token-to-Token method to enrich the tokenization process [29]. Moreover, Liu *et al* designed a swin transformer to include patch fusion and cyclic shift to enlarge the perception of contextual information in tokens [27]. Researches also explored the transformer for low vision task [21, 24, 25]. However, transformer in LDCT denoising has not been well explored. Zhang *et al* designed a TransCT-net to utilize transformer in the high frequency (HF) and low frequency (LF) composite inference [32]. Nevertheless, Zhang's work includes a lot of convolutions in the HF/LF tokenization and Piecewise Reconstruction module. So far, there is no convolution-free model for LDCT denoising. In this paper, for the first time, we propose a convolution-free Transformer Encoder-decoder Dilation (TED-net) model and evaluate its performance compared with other state-of-the-art models.

The rest of this paper is organized as follows. Part II introduces the key methods used in our proposed model. Part III reports the experiment details and results. Part IV discusses some related issues and make a conclusion.

## 2 Methods

In this paper, we propose a convolution-free T2T vision transformer-based Encoder-decoder Dilation network (TED-net). As shown in **Fig. 1**, in the encode part, the model includes Tokenization block, Transformer Block (TB), Cyclic Shift Block (CSB), Token-to-Token block with Dilation (T2TD) and without dilation (T2T). The decoder part includes T2T, T2TD, Inverse Cyclic Shift Block (ICSB) and Detokenization Block.

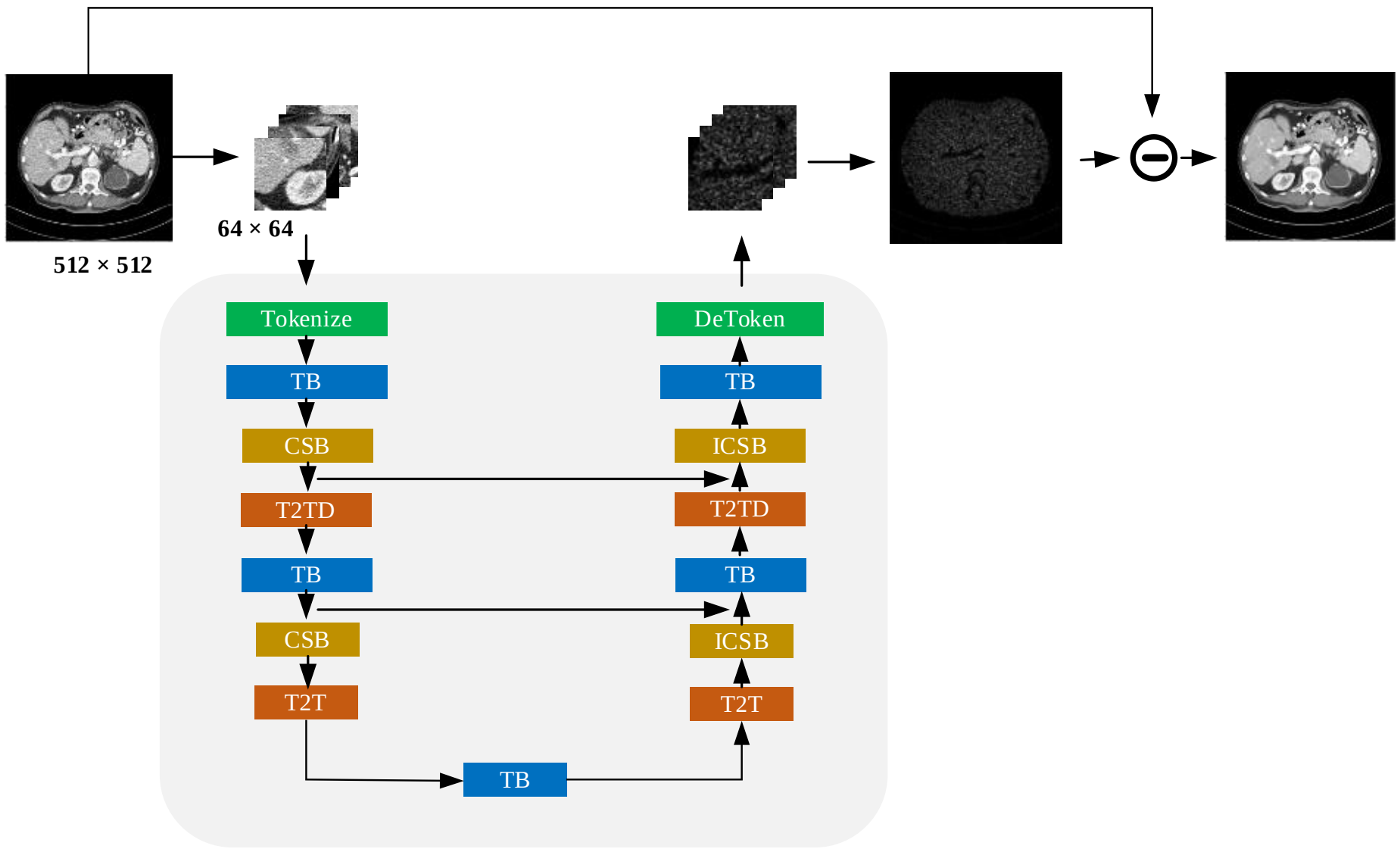

**Fig. 1.** The pipeline of our proposed TED-net. Tokenize block uses unfold to extract tokens from image patches while DeToken block applies Fold to convert tokens back to image. TB includes a typical transformer block. CSB uses cyclic shift operation and ICSB employs inverse cyclic shift. T2T incorporates Token-to-Token block to enhance tokenization while T2TD includes Dilation in the T2T tokenization process. The final image is obtained by subtracting the model residual output from the noisy input image.

## 2.1 Noise Model

The LDCT denoising task attempts to recover a clean NDCT image $y \in \mathbb{R}^{N \times N}$ from a matching noise LDCT image $x \in \mathbb{R}^{N \times N}$ by using a general denoising model $D : \mathbb{R}^{N \times N} \to \mathbb{R}^{N \times N}$, and the Mean-Square Error (MSE) loss is defined as follows:

$$\underset{D}{argmin}\, L_{MSE} = \left\| D(x) - y \right\| . \quad (1)$$

In this paper, we propose a Transformer-based model $T : \mathbb{R}^{N \times N} \to \mathbb{R}^{N \times N}$ to learn the deep features and capture the noise pattern of the image. Then we recover the clean image by combining both the output residual image and the original noisy image.

$$y = T(x) + x . \quad (2)$$

## 2.2 Transformer Block

In the Transformer Block (TB), we utilize a traditional transformer in the encoder and decoder stage between two T2T blocks which contain Multiple head Self-Attention (MSA), Multiple Layer Perceptron (MLP) and residual connection to promote the expressive power of this module. The output of TB $T' \in \mathbb{R}^{b \times n \times d}$ has the same size as the input tokens $T \in \mathbb{R}^{b \times n \times d}$. Here $b$ is batch size, $n$ is the number of tokens, and $d$ is the token embedding dimension.

$$T' = \mathrm{MLP}(\mathrm{MSA}(T))\,. \tag{3}$$

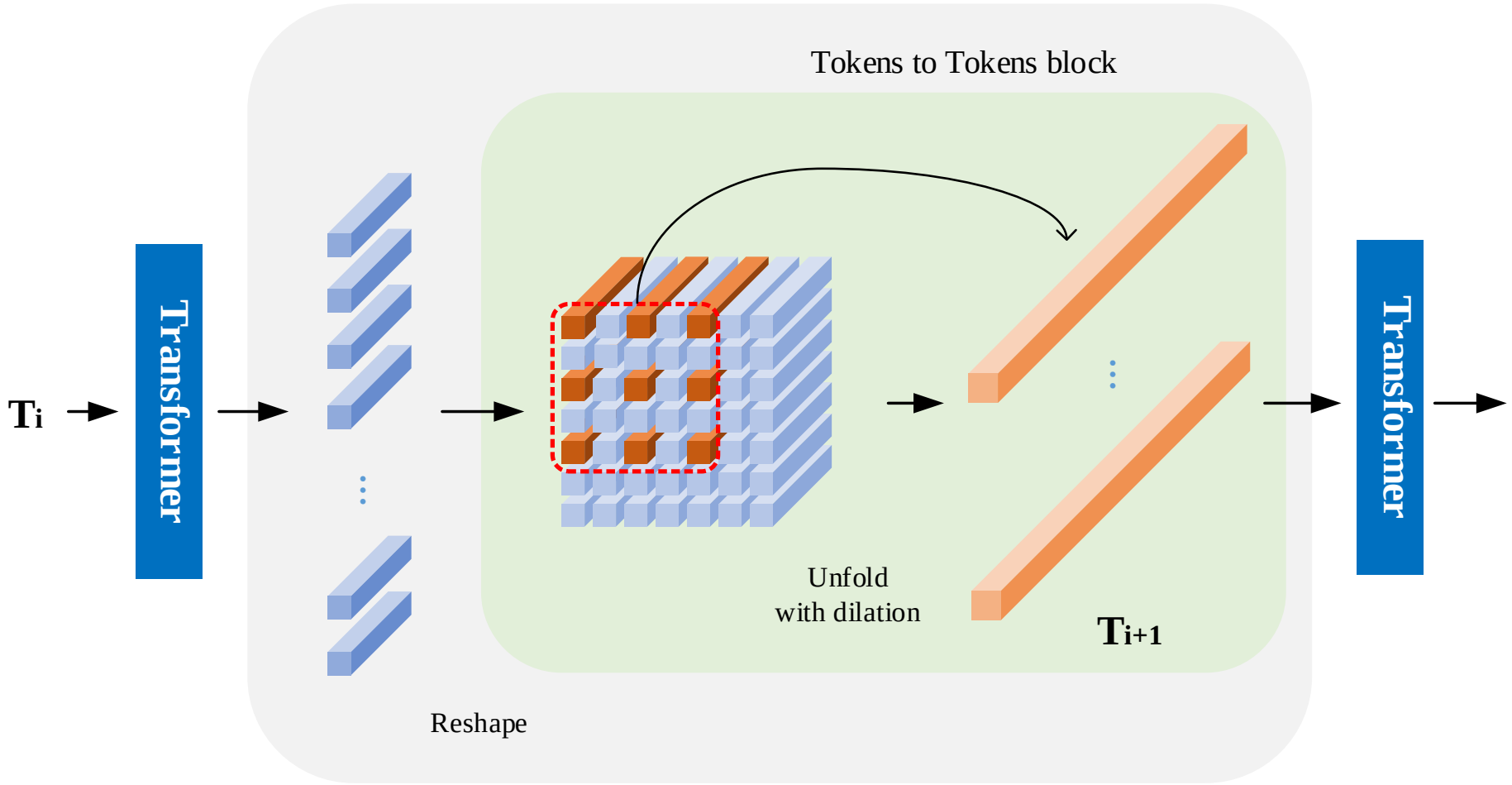

**Fig. 2.** The architecture of T2TD block which includes reshape and unfold with dilation.

### 2.3 Token-to-token Dilation Block

Token-to-Token (T2T) block is recently utilized to overcome the simple tokenization of image in vision transformer. Traditional tokenization only includes one tokenization process using either reshape or convolution, while the T2T block adopts a cascade tokenization procedure. We further use dilation in the tokenization process to refine the contextual information fusion and seek relation across larger regions. **Fig. 2** illustrates the structure of T2TD block which consists of reshape and soft split with dilation.

**Reshape.** Given tokens $T \in \mathbb{R}^{b \times n \times d}$ from last stage, they are first transposed to $T^{T} \in \mathbb{R}^{b \times d \times n}$ and then reshaped into $I \in \mathbb{R}^{b \times c \times h \times w}$.

$$I = \mathrm{Reshape}(T)\,. \tag{4}$$

where $c = d$ and $h = w = \sqrt{n}$ are the channel, height and width of feature map, respectively.

**Soft Split with Dilation.** With the feature maps from the reshape stage, the soft split stage will retokenize the reshaped feature map using unfold operation. In this stage the four dimension feature maps $I \in \mathbb{R}^{b\times c\times h\times w}$ are converted back to three dimensional tokens $T'' \in \mathbb{R}^{b\times n'\times d'}$. Through this operation, the number of tokens are reduced by combining several neighboring tokens into one unit though the embedding dimension is increased accordingly with several tokens concatenated together.

$$T'' = \mathrm{SoftSplit}(I)\,. \tag{5}$$

As demonstrated in **Fig. 2**, dilation is also used in the unfold process to capture the contextual information with longer dependence. After the soft split with dilation, the input feature maps $I \in \mathbb{R}^{b\times c\times h\times w}$ become $T''' \in \mathbb{R}^{b\times n''\times d''}$ where $d'' = c \times \prod kernel$ and the total number of tokens $n''$ after the soft split operation is calculated as:

$$n'' = \left\lfloor \frac{h - dilation \times (kernel - 1) - 1}{Stride} + 1 \right\rfloor. \tag{6}$$

where $dilation, kernel,$ and $stride$ are related parameters in the Unfold operation.

**Cyclic Shift.** After the reshape process in the encoder network, we employ the cyclic-shift to modify the shaped feature map. The pixel values in the feature map are assembled in a different way that will add more information integration in the model. Then, an inverse cyclic shift is performed in the symmetric decoder network to avoid any pixel shifts in the final denoising results. **Fig. 3** exhibits the cyclic shift module and inverse cyclic shift module.

$$I' = \mathrm{CyclicShift}(I)\,. \tag{7}$$

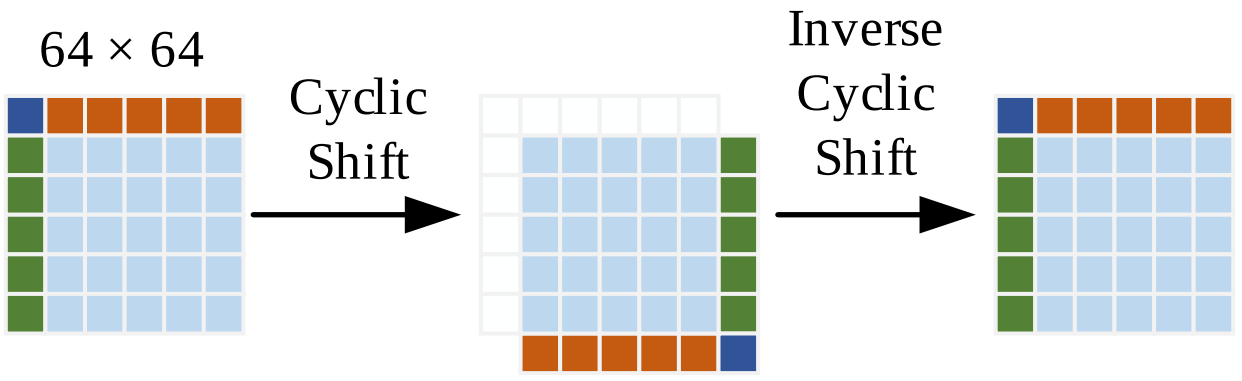

**Fig. 3.** The structures of Cyclic Shift and Inverse Cyclic Shift operations to enrich the tokenization process by fusion different kernel area.

## 3 Experiments and Results

In this part, the data preparation, experiment settings and comparison results are presented. Our model is trained and evaluated on a public dataset, and the results show that our model outperforms other state-of-the-art models.

**Dataset**

A publicly released dataset from 2016 NIH-AAPM-Mayo Clinic LDCT Grand Challenge is used for model training and testing. We employ the patient L506 data for evaluation and the other nine patients for model training. The pairs of quarter-dose LDCT and normal-dose CT (NDCT) images are used to train the model. We randomly extract 4 image patches of 64×64 from each original image of size 512×512 in every epoch. Data augmentation is also applied to enlarge the dataset where we keep a copy of original image and then randomly apply image rotation (90 degrees, 180 degrees or 270 degrees) and flipping (up and down, left and right) to the original one.

**Experiment Settings**

The experiments are running on Ubuntu 18.04.5 LTS, with Intel(R) Core (TM) i9-9920X CPU @ 3.50GHz using PyTorch 1.5.0 and CUDA 10.2.0. The models are trained with 2 NVIDIA 2080TI 11G GPUs. Here are the details of our experiment setting: in the encoder block, our model consists of three soft split stages, two transformer layers, and two cyclic shift layers, while in the decoder block it includes three inverse soft split stages with fold operations, two transformer blocks and two corresponding inverse cyclic shift operations. One additional transformer layer is between the encoder and decoder parts to further incorporate more feature inference. The kernel size for the three unfold/fold operations are 7×7, 3×3, 3×3 with a stride of (2,1,1) and a dilation of (1,2,1), respectively. Moreover, the token dimension in the encoder/decoder part is 256, and pixel quantity is 2 for the two cyclic shift layers. We use a patch number of 4 for training and the epochs are 4000. Adam is adopted to minimize the MSE loss with an initial learning rate of 1e-5. In the evaluation stage, we segment the 512×512 image into overlapped 64×64 patches, and only crop out the center part of the model output to aggregate to the final whole predictions to overcome the boundary artifacts.

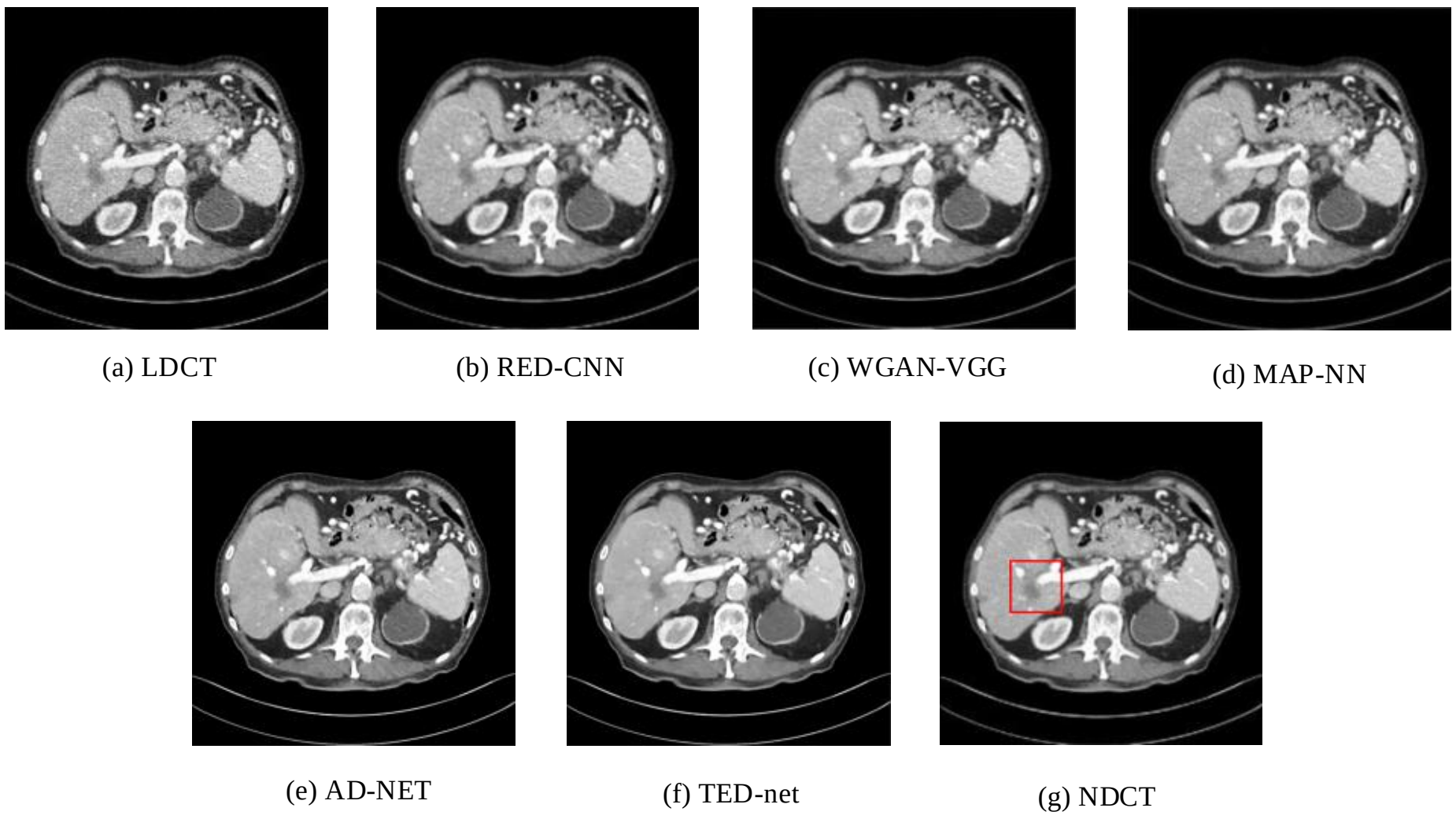

**Fig. 4.** The denoising results of different networks on L506 with lesion No.575. The display window is [-160, 240] HU. (b)-(f) are from RED-CNN, WGAN-VGG, MAP-NN, AD-NET and our proposed TED-net, respectively. (a) and (g) are LDCT and NDCT images.

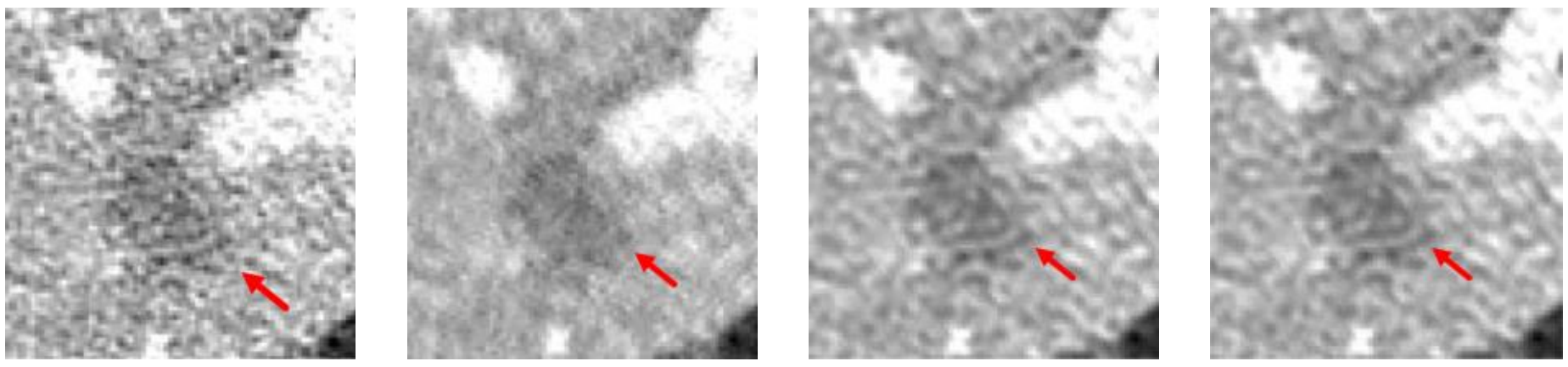

(a) LDCT (b) RED-CNN (c) WGAN-VGG (d) MAP-NN

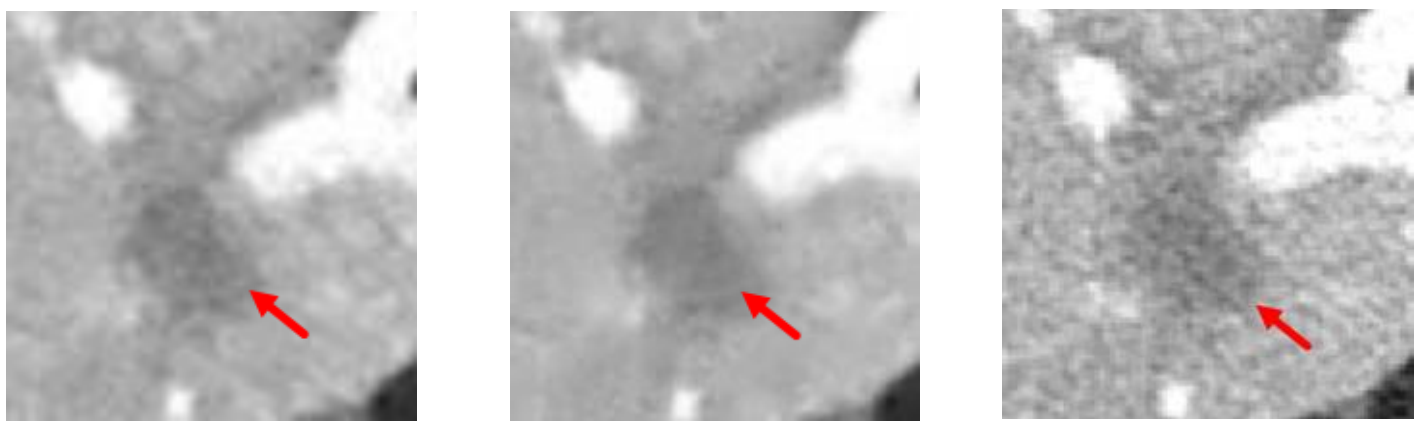

(e) AD-Net (f) TED-net (g) NDCT

**Fig. 5.** The amplified ROIs of different network outputs in the rectangle marked in **Fig. 4**.

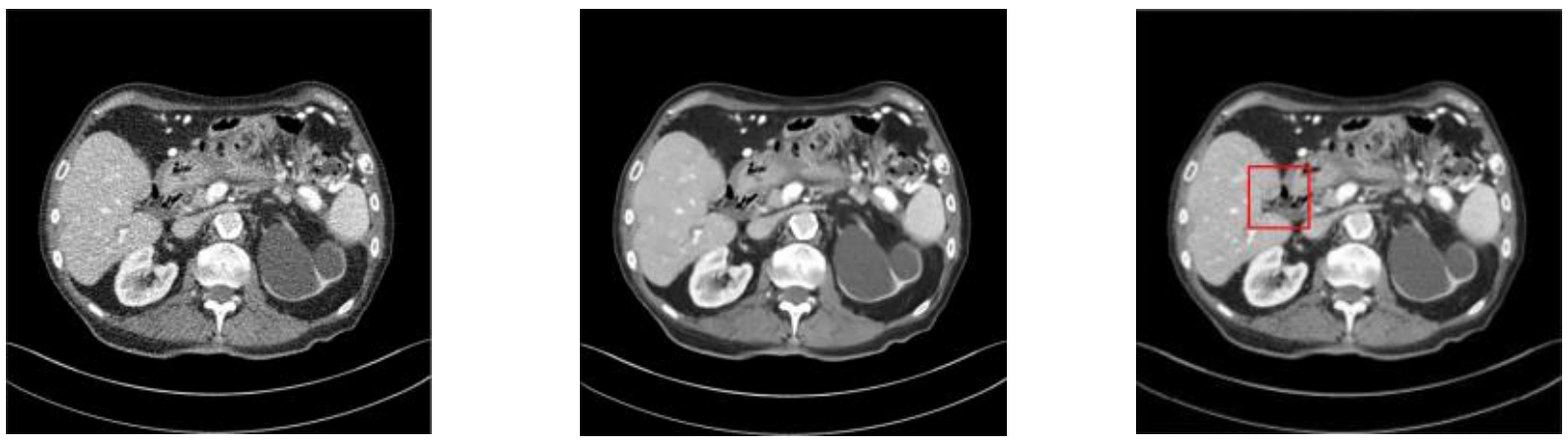

(a) LDCT (b) TED-net (c) NDCT

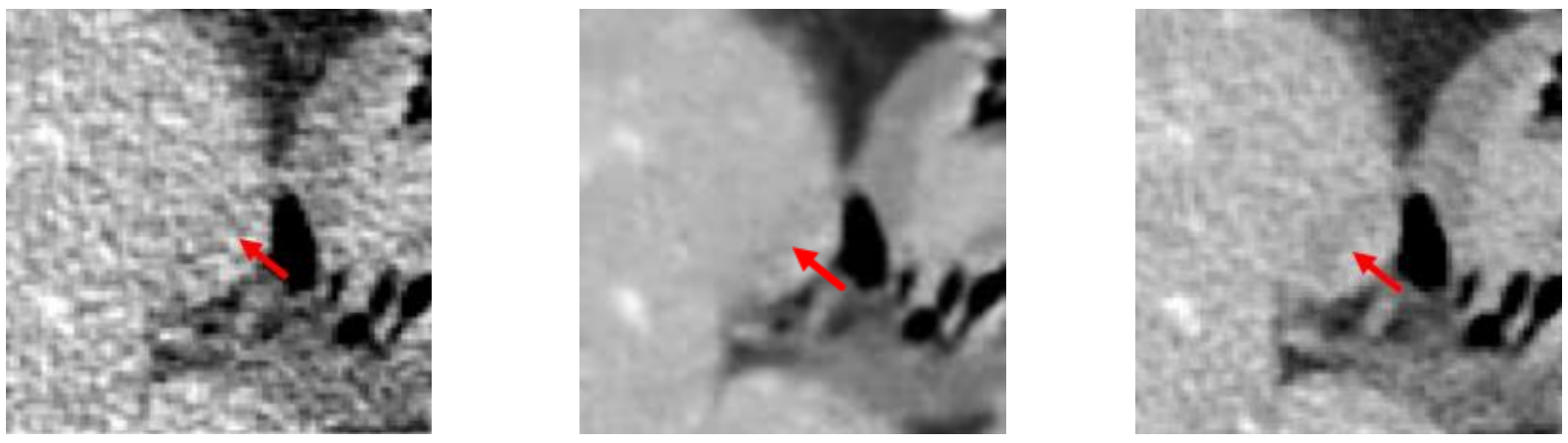

(d) LDCT (e) TED-net (f) NDCT

**Fig. 6.** The performance of TED-net on case L506 with lesion No. 576 and complementary magnified RIOs.

**Comparison results**

SSIM and RMSE are adopted to quantitatively measure the quality of the denoised image. Our model is compared with state-of-the-art baseline algorithms: RED-CNN [11], WGAN-VGG [12], MAP-NN [33], and AD-NET [34]. RED-CNN, MAP-NN, and WGAN-VGG are popular low dose CT denoising models while AD-NET has high performance on gray image denoising. We retrain AD-NET on AAPM dataset with the same setting as other methods and obtain a comparison result. **Fig. 4** shows the results of different networks on L506 with lesion No.575. **Fig. 5** demonstrates the amplified ROIs from the rectangular area marked in **Fig. 4**. **Fig. 6** illustrates the performance of our TED-net on lesion No.576. **Fig. 4**, **Fig. 5** and **Fig. 6** show that our TED-net has a better performance in removing the noise/artifact and maintaining high-level spatial smoothness while keeping the details of the target image. However, other methods have more blotchy noisy textures. Additionally, quantitative results from **Table 1** also confirm that our model outperforms other models.

**Table 1.** Quantitative results of different methods on L506

| Method | SSIM | RMSE |
|---|---|---|
| LDCT | 0.8759 | 14.2416 |
| RED-CNN | 0.8952 | 11.5926 |
| WGAN-VGG | 0.9008 | 11.6370 |
| MAP-NN | 0.8941 | 11.5848 |
| AD-NET | 0.9041 | 9.7166 |
| TED-net* | 0.9144 | 8.7681 |

## 4 Conclusion

In this paper, a novel pure transformer-based convolution-free LDCT denoising algorithm is developed for clinical applications. In contrast, the most state-of-art models are based on CNN. To the best of our knowledge, this is the first research to apply pure transformer for LDCT denoising. Our contributions are mainly three-folds: (1) A convolution-free U-net like T2T-vit-based denoising transformer model is developed. (2) The dilation is used in the T2T stage to enlarge the receptive field to obtain more contextual information from the feature-maps. (3) A cyclic shift is used to furthermore refine the mode of image tokenization. Experimental results show our model outperforms other state-of-the-art models with the highest SSIM value and smallest RMSE value. In the future, this model can be further slimmed with a more powerful tokenization without downgrading of images.